# SMART SUMMARIZER FOR BLIND PEOPLE

Mona teja K[1], Mohan Sai.S[2], H S S S Raviteja D[3], Sai Kushagra P V[3],

[1,2,3,4]UG Student, School of Computer and Engineering, VIT Vellore, India

[1]monateja9999@gmail.com, [2]singamsettymohansai10@gmail.com, [3]ravitejadonthu1@gmail.com, [4]saikushagra@gmail.com

*Abstract*—In today's world, time is a very important resource. In our busy lives, most of us hardly have time to read the complete news so what we have to do is just go through the headlines and satisfy ourselves with that. As a result, we might miss a part of the news or misinterpret the complete thing. The situation is even worse for the people who are visually impaired or have lost their ability to see. The inability of these people to read text has a huge impact on their lives. There are a number of methods for blind people to read text. Braille script in particular is one of the examples, but it is a highly inefficient method as it is really time taking and requires a lot of practice. So, we present a method for visually impaired people based on the sense of sound which is obviously better and more accurate than the sense of touch. This paper deals with an efficient method to summarize a news into important keywords so as to save the efforts to go through the complete text every single time. This paper deals with many API's and modules like tesseract, GTTS and many algorithms have been discussed and implemented in detail such as Luhn's Algorithm, Latent Semantic Analysis Algorithm, Text Ranking Algorithm. And the other functionality that this paper deals with is converting the summarized text to speech so that the system can aid even the blind people.

## I. INTRODUCTION

Giving machines an ability to think has always been a far-fetched dream for humans since ancient times. But since the development of the concept of Machine Learning in the past few decades, giving machines a "brain" is no longer a dream. Yes, the machines today can be made to learn and apply that learning to perform a lot of functions that only humans could do earlier. Some of such applications of Machine Learning are generating the summary of a given piece of text and converting a piece of text to speech. So, it's clear that this paper uses an approach based on the concepts of Machine Learning to brief the summary of a piece of text to deliver the content in very less time and in a clear-cut way. And more over it contributes to different fields and can be attached or implemented in many other systems. Another motive is to give visually impaired people the ability to get the news without anyone's help in a much more efficient manner. In this paper Section II will discuss about the Importance of the proposed system and Section III discuss about the Literature review and Section IV shows the system Architecture. Section V gives the details of API's used and Section VI shows the experiment.

## II. IMPORTANCE OF SYSTEM

The technique that we present in this paper is very useful for visually impaired people as well as the people who are not able to read newspapers because of their tight schedule. It acts as an interface to retrieve the content, whatever it may be, in textual form just by using OCR (Optical Character Recognition) module which runs through image processing. After getting the text, now by applying the Luhn's algorithm the content is summarized into a few important and self-explanatory keywords, thus, saving the efforts to go through the complete news. In the next step, this summarized content is converted to voice that can be directly perceived by for the visually impaired people. Thus, the content can be delivered to them accurately, without anyone's help.

## III. LITERATURE REVIEWS

In paper [1], it mainly comes with two problems, one is searching for the relevant document, and number of such documents available for that particular information. With that scenario it comes with a conclusion of creating a technique which automatically summarizes the text which is very important to overcome the above proposed problems by the paper. This paper tries to compress the content to a smaller information but keeping the meaning of the content alive.

In [2] it talks about the different methods to be followed for decreasing the content by using the centrality on the similarity graph method. It compares the difference between the degree-based methods and the centroid based methods and finally concludes that the degree-based methods are better. The main drawback in this method is that the technique used is insensitive to the noise in the data.

In [3] it shows and proves that the part of the text which are present in the content and which repeat a lot have a high probability to be present on the summarized content or they are the better terms to be present on the summarized content. This paper has coined a term weight on each content (which is the frequency count of that particular text in the content) which give good results when kept in the summarized content.

J MacQueen [4] presents some methods for the classification of multivariate observations, describing a process for the partitioning of N-dimensional population into p sets depending upon the sample. The author explains how the concept of K-Means has to be theoretically interesting apart from being applicable in many practical problems. According to him, K-means Clustering is a specific and one of the most widespread method of clustering. The major part of his research deals with K-Means and some results have been obtained on its asymptotic behavior along with their proofs. Also, several applications and some preliminary results from the conducted experiments to explore the potential inherent in the K-Means algorithm have been discussed.

Another unsupervised approach has been presented by Regina Barzilay and Lillian Lee [6] to paraphrase using Multi-Sequence Alignment, in which they have addressed the text-to-text generation problem of sentence-level paraphrasing that is even more difficult than word- or phrase-level paraphrasing. They present an approach that applies multiple-sequence alignment to sentences gathered from the unannotated corpora. The system is made to learn, a set of paraphrasing patterns and then it automatically determines how to apply those patterns to paraphrase sentences. By using the machine learning concepts, the authors have developed a system that accurately derives paraphrases, thus, outperforming the baseline systems.

David Yarowsky has presented another unsupervised approach for word sense disambiguation [7] that excels in performance. The algorithm, when trained on unannotated English text, rivals supervised techniques that require time-consuming hand annotations, in performance and accuracy. The performance accuracy exceeds 96% as tested by the author when exploited on an iterative bootstrapping procedure. The algorithm avoids the need for any kind of costly hand-tagged training data as it exploits powerful properties of the human language, namely, one sense of collocation and one sense per discourse, to reach the outcome.

## IV. SYSTEM ARCHITECTURE

Figure-I depicts the Architecture of the proposed system.

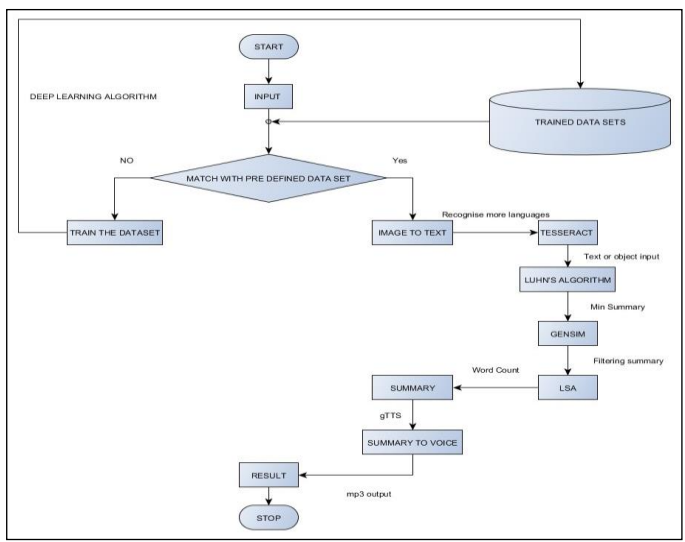

Figure-I

## V. APIs AND MODULES USED

i) The first module that has been used in the system is the TESSERACT: GOOGLE API, which comes under Optical Character Recognition (OCR) engine with support for Unicode and moreover it has the ability to recognise over several languages out of the box. It is flexible and can further be trained for the recognition of other languages too.
This module can be used to detect text on handheld devices and even to identify spam images from Gmail and other common platforms.

ii) The next important module used is the LUHN ALGORITHM. This algorithm is the well-known check sum formula that is used to validate a variety of ID Numbers like IMEI and Credit card numbers. This algorithm is freely available to the common public. The reason for its design was for the protection against accidental errors and not malicious attacks.

iii) Another library we've used is the GENSIM PYTHON LIBRARY, which is also an open source library used for Natural Language Processing (NLP), with specification in Topic Modelling. It can also perform similarity detection and retrieval (IR) and document indexing when provided with large corpora. The target users are the NLP and the IR community.
The task of summarization of the given piece of text is a classic one and has been studied from different perspectives from time to time. We followed the TEST RANKING ALGORITHM for this. The initial step is to pick a subset of a words from the text so that the summary determined by it is as close to the original text as possible. The subset, named the summary, should be logical and understandable. This does not mean that the system determines the most common words only, but the most relevant words available. . The task is not about picking the most common words or entities. We will use the Naive way to perform the neural network training. And moreover, GENSIM is an NLP (Natural Language Processing) algorithm will also be implemented in python.

iv) The next algorithm used is the LATENT SEMANTIC ANALYSIS (LSA) algorithm in python. LAS us a mathematical algorithm that tries to determine Latent Relationships within a collection of the given documents, thus, looking at all the documents as a whole rather than looking at each document separately to identify the relationships. Thus, it determines sets of related words and include the relevant results from the complete set when we search for any word in the set. We have used this module to limit the number of words in the summary generated to keep it as compact as possible.

v) Another important module we have used is the GOOGLE TEXT-TO-SEARCH (GTTS) module. This module alone handles one of the main functionalities of our model, that is, converting the Text we provide to it to Speech for the visually impaired people. By using the Deep Learning algorithms, we can train the datasets, which are then used to recognise the real objects using computer vision.

## VI. EXPERIMENT

In the first phase the Image is given as input to optical character recogniser engine for this we have used Tesseract an open source Google API. And the text which is extracted at an accuracy of 98.72% which is pretty good enough for the proper text detail for the further process to maintain the details of the context. This has been pretrained from the data with lot of handwritten digits and with different fonts in different languages. In this case a sample image is given as input and the text is given as output is shown in fig 2

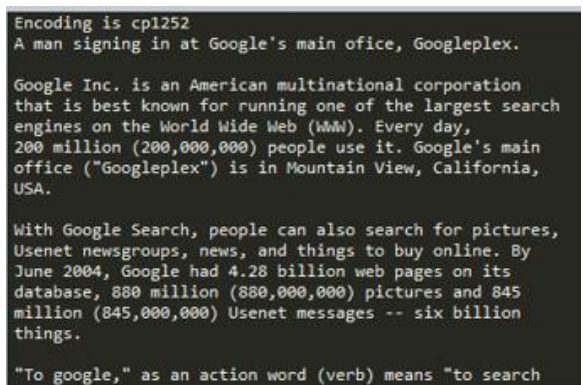

Figure 2(Part of Text extracted from the Image)

In the next step of first phase is there might be a lot of data that might occur from the image and so we have used luhn's algorithm. This large sentence can be summarized using the parameters by finding the intersections of the paragraph and the content has been splitted into sentences and then the formating take place and the which is helpful for finding the ranks of the sentences which will be helpful for the text summarization based on the ranks of the sentences. In this the summary ratio maintfained is 96.512% and the orignal of length found is 4445 words which has been reduces to 155 words and is shown in figure 3.

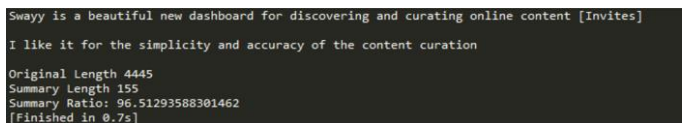

Figure 3(Text summarization)

Then after to maintain the context with the same proper meaning which is python framework for fast Vector space modelling. Genism is used for topic modelling, document indexing and similarity retrieval with copra. The Latent Semantic model is the mathematical model which will search for identical latent relations which will help in summarizing. And next Google API is used for converting the text to speech and which takes in the text document and output an mp3 format which can be simply heard by the user the whole text in summarized format.

The next phase is to recognize the object which is in front of impaired person and to notify the person the object is present. For this object recognition we use a Deep Learning model with the help of computer vision techniques. The dataset used is CIFAR dataset which consists of 1000 categories and this data for the feature extraction HOG (Histogram of oriented gradients) and these fed to fully neural network which is which are 28 layers deep with 3995 neurons fully connected and the final layers with 1000 outputs and the predictions are for recognition we use opencv framework for the object detection and the prediction of the object with accuracy of 95.68% is shown in figure 4.

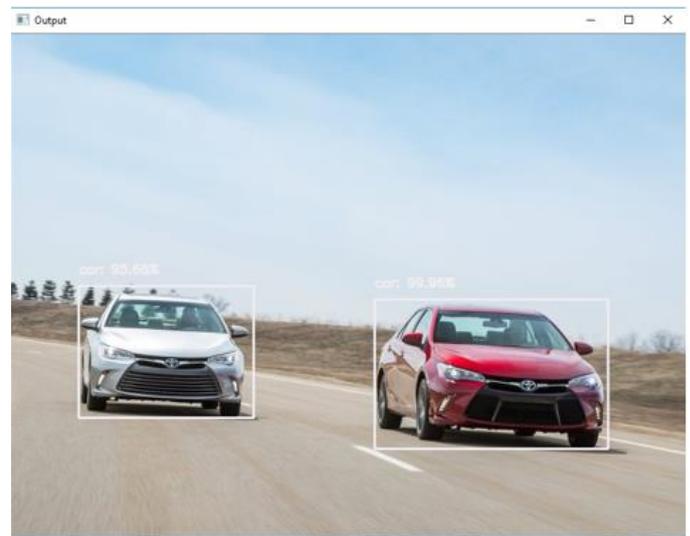

Figure 4(Prediction of object in front of an impaired person)

## VII. CONCLUSION

The way people in the world are approaching to make the world a better place to live in is through optimization and automation. The model we present in this concept-oriented paper can prove to be really useful to the blind people because of the results we obtained. The model summarizes the content presented to it, based on neural network algorithms and that summarized content is finally converted to speech. This brings a major change in lives of the blind people a lot, which is one step success for us. In the end, we were able to extract the basic summary of any piece of text we presented to it. This summary was tested for correctness based on the number of keywords that were obtained as the output and their importance in the text, and we were able to achieve a good efficiency and accuracy, thus, bringing the proposed system to the expected conclusion.

## VIII. FUTURE SCOPE

The paper we have done, we want to convert into a matured project by integrating the things we have done in the form of a mobile application. Later we want to integrate with the google search, so that if a person searches in the google with a particular keyword, it gives the summarized content of the pages based on the keyword and then the summarized content will be generated to voice, so that even blind people will be part of google search.

REFERENCES

[1] Text Summarization:An Overview, By Samrat Babar

[2] Erkan ,G. & Radev, D. (2004). LexRank: Graph-based Lexical Centrality as Salience in Text Summarization. Journal of Artificial Intelligence Research 22 (2004) 457-479

[3] ] Villatoro-Tello, E., Villaseñor-Pineda, L., Montes-y-Gómez, M.: Using Word Sequences for Text Summarization. In: Sojka, P., Kopeček, I., Pala, K. (eds.) TSD 2006. LNCS (LNAI), vol. 4188, pp. 293–300. Springer, Heidelberg (2006)
Mahajan, M.; Nimbhorkar, P.; Varadarajan, K. (2009). "The Planar kMeans Problem is NP-Hard". Lecture Notes in Computer Science 5431: 274–285

[4] J. B. MacQueen (1967): "Some Methods for classification and Analysis of Multivariate Observations, Proceedings of 5-th Berkeley Symposium on Mathematical Statistics and Probability", Berkeley, University of California Press, 1:281-297

[5] Kaliappan, J., Shreyansh, J., & Singamsetti, M. S. (2019, March). Surveillance Camera using Face Recognition for automatic Attendance feeder and Energy conservation in classroom. In *2019 International Conference on Vision Towards Emerging Trends in Communication and Networking (ViTECoN)* (pp. 1-5). IEEE.

[6] Sai, S.M., Gopichand, G., Reddy, C.V. and Teja, K., 2019. High Accurate Unhealthy Leaf Detection. *arXiv preprint arXiv:1908.09003*.

[7] Sai, S. M., Muppa, S. K., Teja, K. M., & Natrajan, P. (2018, December). Advanced Image Processing techniques based model for Brain Tumour detection. In *2018 4th International Conference on Computing Communication and Automation (ICCCA)* (pp. 1-6). IEEE.

[8] H. Wu and R. Luk and K. Wong and K. Kwok. "Interpreting TF-IDF term weights as making relevance decisions". ACM Transactions on Information Systems, 26 (3). 2008.

[9] Barzilay, R., & Lee, L. (2003). Learning to paraphrase: An unsupervised approach using multiple-sequence alignment. In Proceedings of HLTNAACL

[10] Yarowsky, D. (1995). Unsupervised word sense disambiguation rivaling supervised methods. In Proceedings of the 33rd Annual Meeting of the Association for Computational Linguistics.

[11] Sai, S. M., Naresh, K., RajKumar, S., Ganesh, M. S., Sai, L., & Nav, A. (2018, April). An Infrared Image Detecting System Model to Monitor Human with Weapon for Controlling Smuggling of Sandalwood Trees. In *2018 Second International Conference on Inventive Communication and Computational Technologies (ICICCT)* (pp. 962-968). IEEE.

[12] Mihalcea, R., Tarau, P.: TextRank: Bringing Order into Texts. In: Proc. Empirical Methods in Natural Language Processing (EMNLP 2004), Barcelona, Spain (2004)

[13] Sidorov, G., Gelbukh, A.: Automatic Detection of Semantically Primitive Words Using Their Reachability in an Explanatory Dictionary. In: Proc. Int. Workshop on Natural Language Processing and Knowledge Engineering, NLPKE 2001, USA, pp. 1683–1687 (2001)